\documentclass[aps,prl,twocolumn,showpacs]{revtex4}

\usepackage{graphicx}
\usepackage{dcolumn}
\usepackage{amsmath}
\usepackage{amssymb}
\usepackage{wasysym}
\usepackage{color}
\begin{document}

\title{Fermi surface reconstruction from bilayer charge ordering in the underdoped high temperature superconductor YBa$_2$Cu$_3$O$_{6+x}$}

\author{N.~Harrison$^1$
}
\author{S.~E.~Sebastian$^2$}

\affiliation{$^1$National High Magnetic Field Laboratory, Los Alamos National Laboratory, Los Alamos, NM 87545\\
$^2$Cavendish Laboratory, Cambridge University, JJ Thomson Avenue, Cambridge CB3~OHE, U.K
}
\date{\today}

\begin{abstract}
We show that a Fermi surface in underdoped YBa$_2$Cu$_3$O$_{\rm 6+x}$ yielding the distribution of quantum oscillation frequencies observed over a broad range of magnetic field can be reconciled with the wavevectors of charge modulations found in nuclear magnetic resonance and  x-ray scattering experiments within a model of biaxial charge ordering occurring in a bilayer CuO$_2$ planar system. Bilayer coupling introduces the possibility of different period modulations and quantum oscillation frequencies corresponding to each of the bonding and antibonding bands, which can be reconciled with recent experimental observations.
\end{abstract}
\maketitle

Our level of understanding of metals typically depends on the extent to which Fermi surface cross-sections observed in quantum oscillation experiments can be explained by comparison with band structure calculations, and the degree to which Fermi liquid behavior prevails~\cite{ashcroft1,shoenberg1,bergemann1,kartsovnik1}. Additional factors, such as a periodic modulation of the electronic states $-$ caused, for example, by a spin- or charge-density wave $-$ can introduce additional gaps, in which case the Fermi surface needs to be understood in terms of the original band structure translated by ordering wavevectors~\cite{kartsovnik1,fleming1,fawcett1,johannes1}. 

In the case of high temperature superconductors, a determination of the electronic structure is essential for understanding the origin of pairing. Yet, the small Fermi surface pocket observed in the underdoped cuprates YBa$_2$Cu$_3$O$_{6+x}$~\cite{doiron1} and YBa$_2$Cu$_4$O$_8$~\cite{yelland1} cannot readily be explained in terms of the unmodified band structure~\cite{chang1,hossain1,sebastian2,sebastian3,yelland1,leboeuf1, sebastian4,vignolle1,sebastian5}. The momentum-space cross-section of  the experimentally observed pocket is roughly thirty times smaller than the size of the large hole Fermi surface sections expected for the CuO$_2$ planes from band structure calculations~\cite{andersen1,elfimov1,carrington1}, with the cyclotron motion having apparently the opposite sense of rotation to that expected~\cite{leboeuf1,chang1}.  A small pocket has  been proposed to originate from  band structure involving the mixing of the chain states with BaO orbitals in YBa$_2$Cu$_3$O$_{6+x}$~\cite{elfimov1,carrington1}, but the relevant band is reported to lie significantly ($\approx$~0.6~eV) below the Fermi energy in photoemission experiments~\cite{hossain1}.

The emergence of long range charge order reported in underdoped YBa$_2$Cu$_3$O$_{6+x}$ by nuclear magnetic resonance (NMR)~\cite{wu1} and x-ray scattering~\cite{ghiringhelli1,chang2,achkar1} may contribute to the observed difference between the small Fermi surface pocket measured and the large Fermi surface expected from band structure. A direct link, however, has yet to be established between the structure of the measured order and reconstruction of the large Fermi surface to yield the observed small pocket. Furthermore, some differences are reported in the ordering wavevectors seen in different experiments~\cite{wu1,ghiringhelli1,chang2,achkar1}, and it remains to be settled whether the observed charge order in YBa$_2$Cu$_3$O$_{6+x}$ is stripe-like~\cite{millis1,yao1,laliberte1} as opposed to biaxial~\cite{harrison1,sebastian3} in nature.

A satisfactory electronic structure model of underdoped YBa$_2$Cu$_3$O$_{6+x}$ would be expected to explain the ordering wavevectors found in both NMR and x-ray experiments, while accounting for the measured distribution of bilayer-split frequencies in quantum oscillation experiments~\cite{audouard1,sebastian2}, as well as other experimental observations relating to the electronic structure~\cite{hossain1,sebastian1,sebastian3,leboeuf1,chang1,riggs1,heatnote}. A single layer scheme of Fermi surface reconstruction by biaxial charge order has previously been shown to yield nodal electron pockets~\cite{harrison1,harrison2,sebastian4,sebastian3} consistent with experimental observations of the small size of the electronic heat capacity at high magnetic fields~\cite{riggs1,heatnote}, the negative electrical transport coefficients~\cite{leboeuf1,chang1} and the predominantly nodal Fermi surface location inferred from photoemission, Fermi velocity and chemical potential oscillation measurements~\cite{hossain1,sebastian1,sebastian3}. In this paper, we show that the introduction of bilayer coupling~\cite{garcia1} to the single layer scheme~\cite{harrison1,harrison2,sebastian4,sebastian3} leads to the possibility of a modulation period and Fermi surface pocket of different size corresponding to each of the bonding and antibonding bands, consistent both with experimental measurements of the charge ordering~\cite{wu1,ghiringhelli1,chang2,achkar1}, and quantum oscillations~\cite{audouard1,sebastian2}.

NMR experiments provide  evidence for long range charge ordering in a magnetic field $B\gtrapprox$~15~T~\cite{wu1} in underdoped YBa$_2$Cu$_3$O$_{6+x}$, while recent x-ray scattering experiments reveal ordering vectors that are two-dimensional~\cite{ghiringhelli1,chang2,achkar1}. The x-ray results are significant for two reasons. First, they reveal the scattering intensity and correlation lengths to be equivalent along the $a$ and $b$ lattice directions of the orthorhombic unit cell~\cite{andersen1}, with the orthorhombicity and CuO chain oxygen ordering having no apparent effect in causing stripe domains to form preferentially along either $a$ or $b$. Second, they reveal the charge ordering to be incommensurate with wavevectors ${\bf Q}_a=(\frac{2\pi}{\lambda_a a},0,Q_z)$ and ${\bf Q}_b=(0,\frac{2\pi}{\lambda_b b},Q_z)$ [of in-plane period $\lambda_a=\lambda_b\approx$~3.2 lattice constants] consistent with a wavevector nesting the bilayer-split bonding band~\cite{ghiringhelli1,chang2,achkar1} (where $Q_z$ refers to the $c$-axis component of the ordering vector). The periods reported in x-ray scattering experiments, however, differ from that ($\lambda_a=$~4.0) reported in NMR studies~\cite{wu1} of similar YBa$_2$Cu$_3$O$_{6+x}$ samples and that ($\lambda\approx$~4.2)~\cite{wilkins1} typically found by x-ray diffraction in the same region of the phase diagram in single layer spin-stripe ordered cuprates~\cite{tranquada1}. Furthermore, no obvious relationship is found between the charge ordering wave vectors found in x-ray experiments on YBa$_2$Cu$_3$O$_{6+x}$ and low energy spin excitation wave vectors~\cite{dai1}.

\begin{figure}
\centering 
\includegraphics*[width=.3\textwidth]{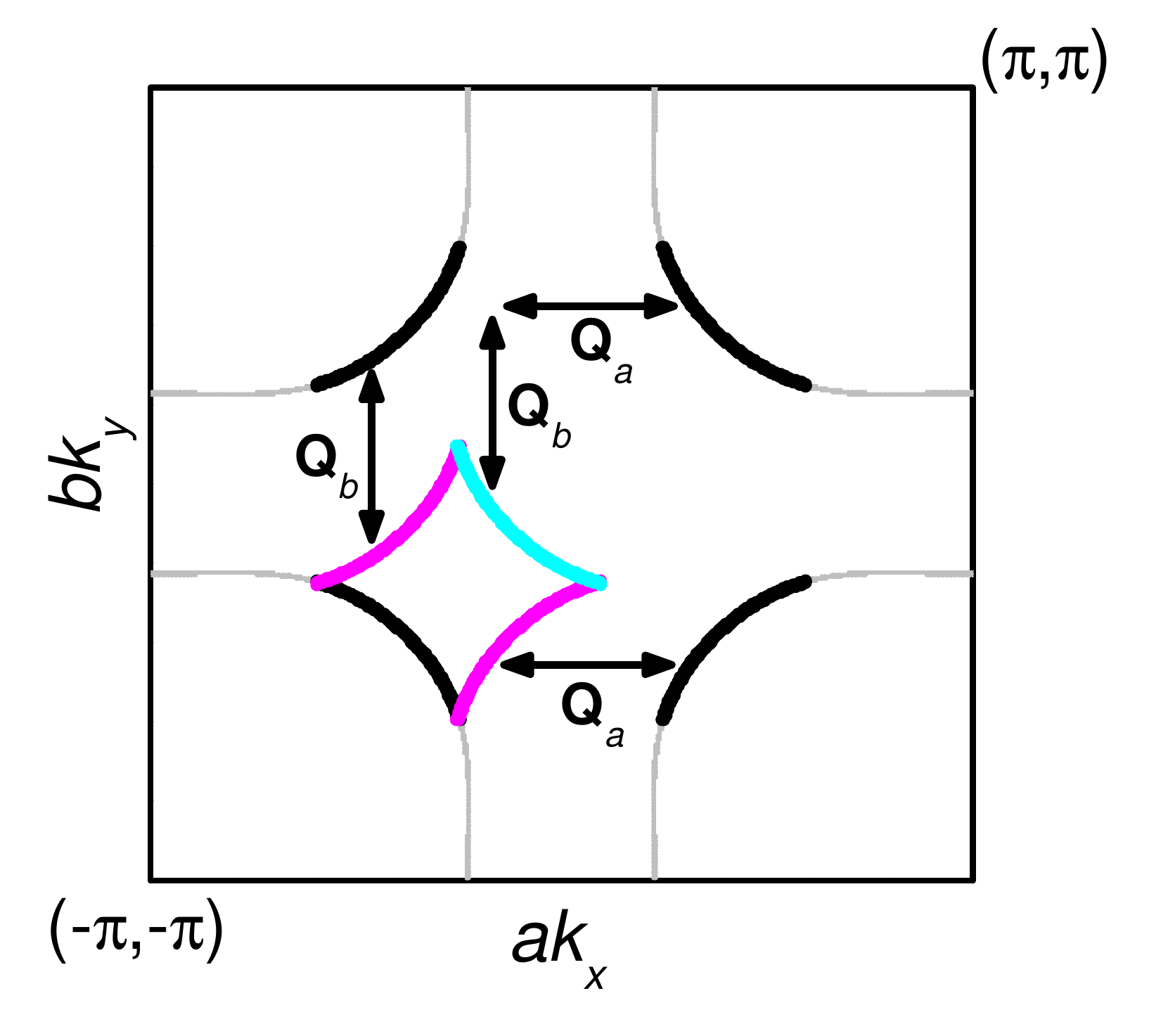}
\caption{Single layer schematic showing how  gapping of the electronic density of states at the antinodes (depicted in grey) leaves remaining arc-like segments of Fermi surface~\cite{harrison1,harrison2,sebastian3} (depicted in black), whose reconnection upon translation by ${\bf Q}_a$ and ${\bf Q}_b$ leads to the possibility of a nodal pocket for which the sense of cyclotron motion in a magnetic field is opposite to that of the original hole sheets. Thick black lines represent $\varepsilon({\bf k})$, thick magenta lines represent $\varepsilon({\bf k}+{\bf Q}_a)$ and $\varepsilon({\bf k}+{\bf Q}_b)$, while thick cyan lines represent $\varepsilon({\bf k}+{\bf Q}_a+{\bf Q}_b)$. No direct coupling occurs between $\varepsilon({\bf k})$ and $\varepsilon({\bf k}+{\bf Q}_a+{\bf Q}_b)$ in the model, however, $\varepsilon({\bf k})$ and $\varepsilon({\bf k}+{\bf Q}_a+{\bf Q}_b)$ are both coupled to $\varepsilon({\bf k}+{\bf Q}_a)$ and $\varepsilon({\bf k}+{\bf Q}_b)$, thus leading to a closed pocket. }
\label{schematic}
\end{figure}

Given the unambiguous planar origin of the two-dimensional charge ordering found in x-ray scattering experiments in YBa$_2$Cu$_3$O$_{6+x}$~\cite{achkar1},  we focus here on a model dispersion solely for the CuO$_2$ bilayers. While photoemission experiments have yet to identify signatures of band folding consistent with charge ordering,  the wave vectors found using x-rays in YBa$_2$Cu$_3$O$_{6+x}$~\cite{ghiringhelli1,chang2}, as well as those found earlier in doping-dependent scanning tunneling microscopy measurements in other cuprates~\cite{wise1,shen1},  suggest Fermi surface nesting at the antinodes~\cite{harrison2,wise1}. The opening of an antinodal gap (shaded light grey in Fig.~\ref{schematic}) that is expected to accompany ordering at such wave vectors is consistent with a reconstructed Fermi surface that involves relative translations of the remaining arc-like segments of Fermi surface (represented by thick black lines in Fig.~\ref{schematic}~\cite{sebastian3,harrison1,harrison2}). These translated segments of Fermi surface would occupy the same regions in momentum space where `Fermi arcs' are observed in photoemission experiments in underdoped cuprates~\cite{hossain1,fournier1,kanigel1,shen1}.

\begin{figure}
\centering 
\includegraphics*[width=.46\textwidth]{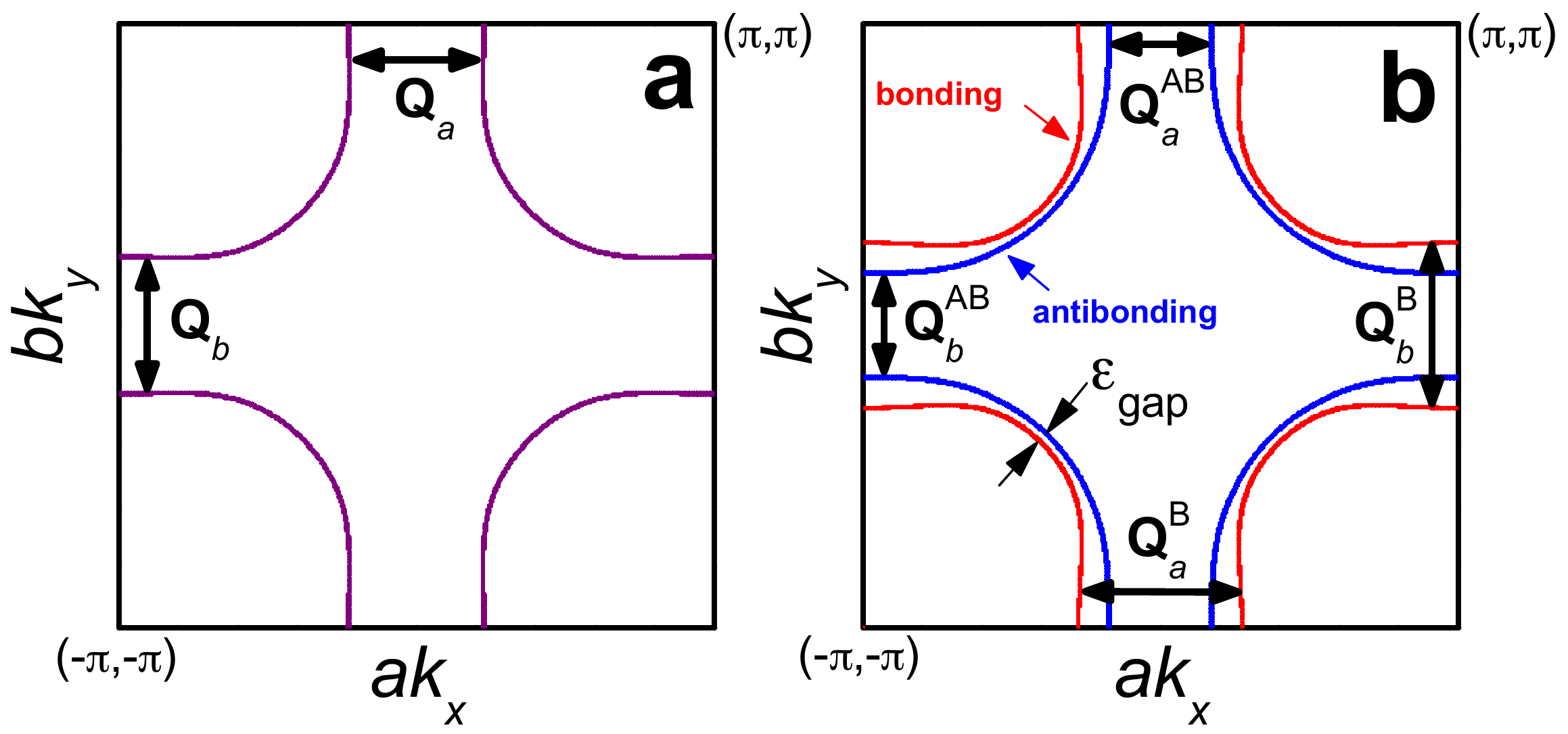}
\caption{(a) Single layer Fermi surface corresponding to $\varepsilon({\bf k})$ as defined in the text. (b) Bilayer Fermi surface after Eq. (\ref{bondinganti}) for $k_z=$~0, where the red and blue lines correspond to bonding and antibonding bands respectively. The cross-sectional area of the Fermi surface represents an effective hole doping of $p=$~8~\%~relative to the half filled band, which we estimate from the total width of the chain Fermi surface seen in photoemission experiments on ortho-II ordered YBa$_2$Cu$_3$O$_{6+x}$~\cite{hossain1}. ${\bf Q}^{\rm B}$ and ${\bf Q}^{\rm AB}$ refer to nesting wavevectors of the bonding and antibonding bands respectively, while ${\bf Q}_a$ and ${\bf Q}_b$ pertain to the different lattice directions in the orthorhombic unit cell~\cite{andersen1}. $\varepsilon_{\rm gap}$ refers to the nodal energy gap.}
\label{Fermisurface}
\end{figure}

The single layer unreconstructed Fermi surface produced using $\varepsilon({\bf k})=\varepsilon_0+2t_{10}[\cos ak_x+\cos bk_y]+2t_{11}[\cos(ak_x+bk_y)+\cos(ak_x-bk_y)]+2t_{20}[\cos 2ak_x+\cos 2bk_y]$ has a single nesting wavevector spanning the antinodal regions [i.e. $|{\bf Q}_a|=|{\bf Q}_b|$ in Fig.~\ref{Fermisurface}(a)]. On introducing bilayer splitting, two nesting vectors now span the bonding and antibonding bands such that $|{\bf Q}_a^{\rm B}|=|{\bf Q}_b^{\rm B}|$ and $|{\bf Q}_a^{\rm AB}|=|{\bf Q}_b^{\rm AB}|$ are inequivalent [shown in Fig.~\ref{Fermisurface}(b)]. For the single-layer dispersion~\cite{choice}, we choose $t_{11}/t_{10}=-$~0.32 and $t_{20}/t_{10}=$~0.16 after Andersen {\it et al.}~\cite{andersen1} and Millis \& Norman~\cite{millis1}. The bonding and antibonding bands are obtained by diagonalizing 
\begin{equation}\label{matrix1} {\bf H}_{\rm bi}=\left( \begin{array}{cc}\varepsilon({\bf k}) & -t_\perp({\bf k})-t_ce^{+ick_z}\\-t_\perp({\bf k})-t_ce^{-ick_z} & \varepsilon({\bf k})\\ \end{array} \right)
\end{equation}
after Garcia-Aldea and Chakravarty~\cite{garcia1}, where
\begin{eqnarray}
t_\perp({\bf k})=\frac{t_{\perp0}}{4}[\cos(ak_x)-\cos(bk_y)]^2\nonumber
\label{inplanebilayer}
\end{eqnarray}
and $t_c$ represent intrabilayer and interbilayer couplings respectively. On evaluating Eq.~(\ref{matrix1}) one obtains
\begin{equation}\label{bondinganti}
\varepsilon({\bf k})^{\rm B,AB}=\varepsilon({\bf k})\mp\sqrt{t_\perp({\bf k})^2+t_c^2+2t_\perp t_c\cos ck_z}
\end{equation}
for the bonding and antibonding bands (shown for $\cos ck_z=0$ in Fig.~\ref{Fermisurface}b), where $c$ refers to the $c$-axis lattice parameter. Setting $t_{\perp0}/t_{10}\approx$~0.5 and $t_c/t_{\perp0}=$~0.25 produces a bilayer splitting at the antinodes [${\bf k}_{\rm antinode}=(\pm\frac{\pi}{a},0,k_z)$ and $(0,\pm\frac{\pi}{a},k_z)$] resembling that in bandstructure calculations~\cite{elfimov1,carrington1} in YBa$_2$Cu$_3$O$_{6+x}$, but with a reduced nodal gap ($\varepsilon_{\rm g}\approx2t_c$) that falls within the experimental upper bound from photoemission~\cite{fournier1} and magnetic quantum oscillation measurements~\cite{sebastian2}.

\begin{figure}
\centering 
\includegraphics*[width=.4\textwidth]{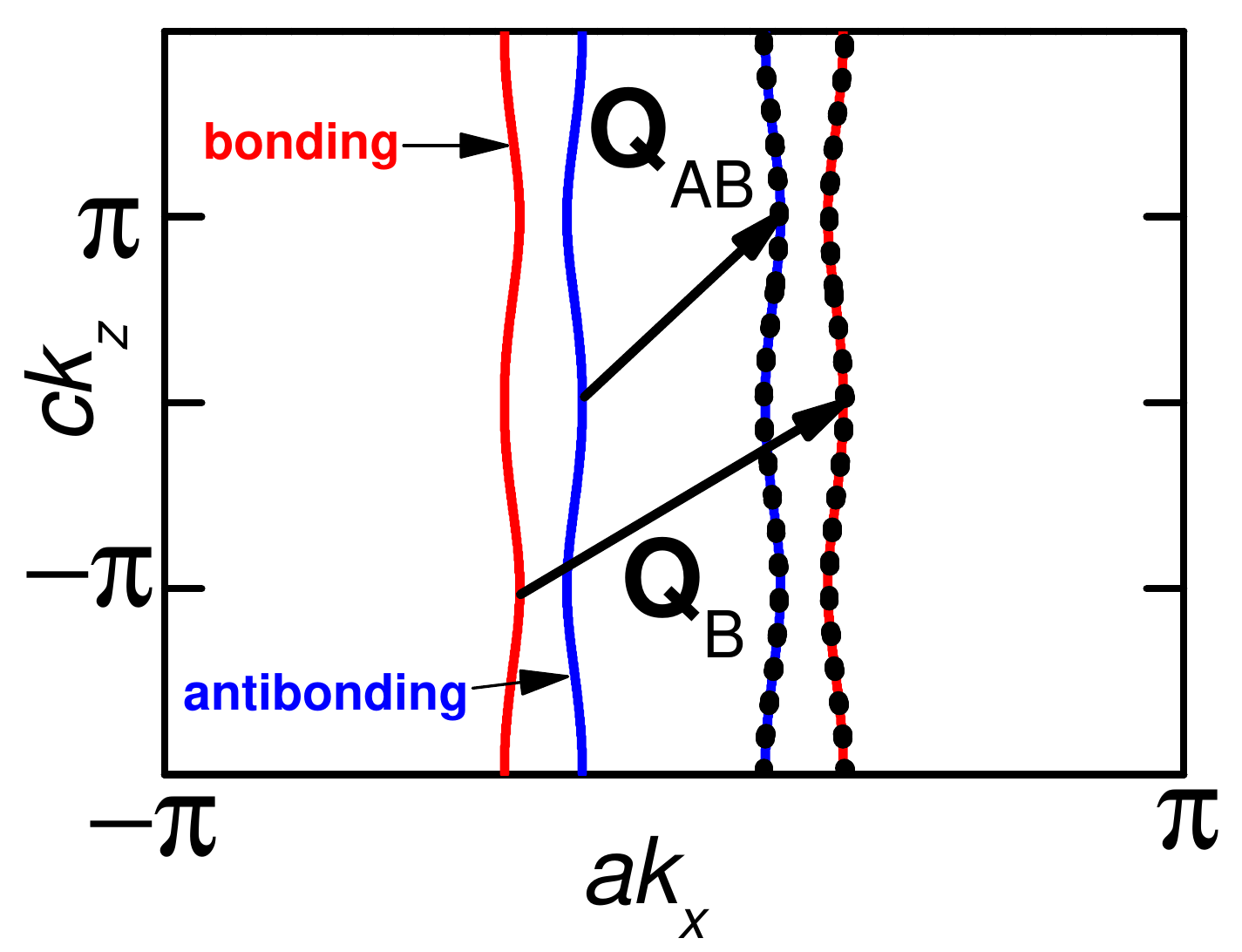}
\caption{Schematic of the interlayer Fermi surface corrugation at $bk_y=\pm\pi$ according to Eqn (\ref{bondinganti}). Owing to a $Q_z=\pi$ component, translation (dotted lines) of the bonding and antibonding Fermi surfaces by ${\bf Q}^{\rm B}$ and ${\bf Q}^{\rm AB}$, respectively, leads to almost perfect interlayer Fermi surface nesting at the antinodes.}
\label{caxis}
\end{figure}

The corrugation of the unreconstructed bilayer Fermi surface with respect to the interlayer momentum $k_z$ (shown in Fig.~\ref{caxis}) caused by $t_{\perp0}$ and $t_c$ implies that Fermi surface nesting is optimized by acquiring a $Q_z=\pi$ component to the wave vector. Such a component produces a reconstructed Fermi surface in which the fundamental component of the corrugation is suppressed, leaving a small residual second harmonic corrugation of order $t_c^2/t_{\perp0}$ [originating from the square root in Eqn.~(\ref{bondinganti})]. Hard x-ray scattering experiments find direct evidence for an interlayer component to the ordering vector~\cite{chang1}. Furthermore, quantum oscillation experiments performed over a broad range of magnetic field and angles are found to be consistent with a bilayer-split Fermi surface that exhibits a comparatively weak corrugation~\cite{sebastian2,sebastian4}. The strong suppression of the interlayer optical conductivity within the underdoped regime~\cite{tranquada3} could be a further signature of Fermi surface corrugation suppressed by nesting. The small size of the residual corrugation term (which we estimate later in the manuscript) compared to the other energy scales enables us to consider a simplified two-dimensional approximation to the Fermi surface, which we obtain by neglecting the $k_z$-dependence in Eqn~(\ref{bondinganti}).

To obtain the full set of reconstructed bands, the relevant couplings ($V_{a,b}^{\rm B,AB}$) between bands with relative translations by $\pm{\bf Q}_a^{\rm B}$, $\pm{\bf Q}_b^{\rm B}$, $\pm{\bf Q}_a^{\rm AB}$ and $\pm{\bf Q}_b^{\rm AB}$ need to be considered [including both the bonding and antibonding bands given by Eqn (\ref{bondinganti})], requiring the construction of a Hamiltonian in the form of a matrix (see e.g. Refs.~\cite{millis1,garcia1}).
Orthogonality between the Bloch wavefunctions of the bonding and antibonding bands ensures that charge-density modulations form independently for those two bands, with the coupling between them being a second order effect~\cite{neglect}. The strongest coupling $V_{a,b}({\bf k}^\prime)$ is therefore expected to occur between states within each of the bands translated by wavevectors intrinsic to that band $-$ i.e. between $\varepsilon({\bf k})^{\rm B}$ and $\varepsilon({\bf k}\pm{\bf Q}^{\rm B}_{a,b})^{\rm B}$ and between $\varepsilon({\bf k})^{\rm AB}$ and $\varepsilon({\bf k}\pm{\bf Q}^{\rm AB}_{a,b})^{\rm AB}$. 

While the complete set of bands for strictly incommensurate order (in which $\lambda^{\rm B}$ and $\lambda^{\rm AB}$ are irrational) requires an infinite number of translations and a matrix of infinite rank, which cannot be diagonalized, it was shown in Ref.~\cite{harrison2} that one can approximate incommensurate behavior by considering rational values of $\lambda$ (in which $\lambda$ is the ratio of two integers) $-$ yielding large but manageable matrices. Provided $0\ll|V_{a,b}({\bf k}^\prime)|<|t_{10}|$, the Fermi surface consists of multiple repetitions of the same electron pocket throughout the Brillouin zone~\cite{harrison2}. Here we go a step further in simplification by considering only the minimal number of terms necessary to produce a single instance of the electron pocket for each band (e.g. those illustrated in Fig.~\ref{schematic}). Hence,
\begin{widetext}
\begin{equation}\label{matrix2}
H^{\rm B,AB}=
\left( \begin{array}{cccc}
\varepsilon({\bf k})^{\rm B,AB} & V_a({\bf k})^{\rm B,AB} & V_b({\bf k})^{\rm B,AB} & 0\\
V_a({\bf k})^{\rm B,AB} & \varepsilon({\bf k}+{\bf Q}_a^{\rm B,AB})^{\rm B,AB} & 0 & V_b({\bf k}+{\bf Q}_a^{\rm B,AB})^{\rm B,AB}\\
V_b({\bf k})^{\rm B,AB} & 0 & \varepsilon({\bf k}+{\bf Q}_b^{\rm B,AB})^{\rm B,AB} & V_a({\bf k}+{\bf Q}_b^{\rm B,AB})^{\rm B,AB}\\
0 & V_b({\bf k}+{\bf Q}_a^{\rm B,AB})^{\rm B,AB} & V_a({\bf k}+{\bf Q}_b^{\rm B,AB})^{\rm B,AB} & \varepsilon({\bf k}+{\bf Q}_a^{\rm B,AB}+{\bf Q}_b^{\rm B,AB})^{\rm B,AB}\end{array} \right),
\end{equation}
\end{widetext}
which must be diagonalized for both bonding and antibonding bands. With these simplifications, we can estimate the size of the pockets, their effective masses and contributions to the electronic density of states for arbitrary values of  $\lambda^{\rm B}$ or $\lambda^{\rm AB}$ $-$ bearing in mind that the complete Fermi surface will have these same pockets repeated by $\pm{\bf Q}_a^{\rm B}$, $\pm{\bf Q}_b^{\rm B}$, $\pm{\bf Q}_a^{\rm AB}$ and $\pm{\bf Q}_b^{\rm AB}$ and combinations thereof throughout the Brillouin zone.
\begin{figure}
\centering 
\includegraphics*[width=.46\textwidth]{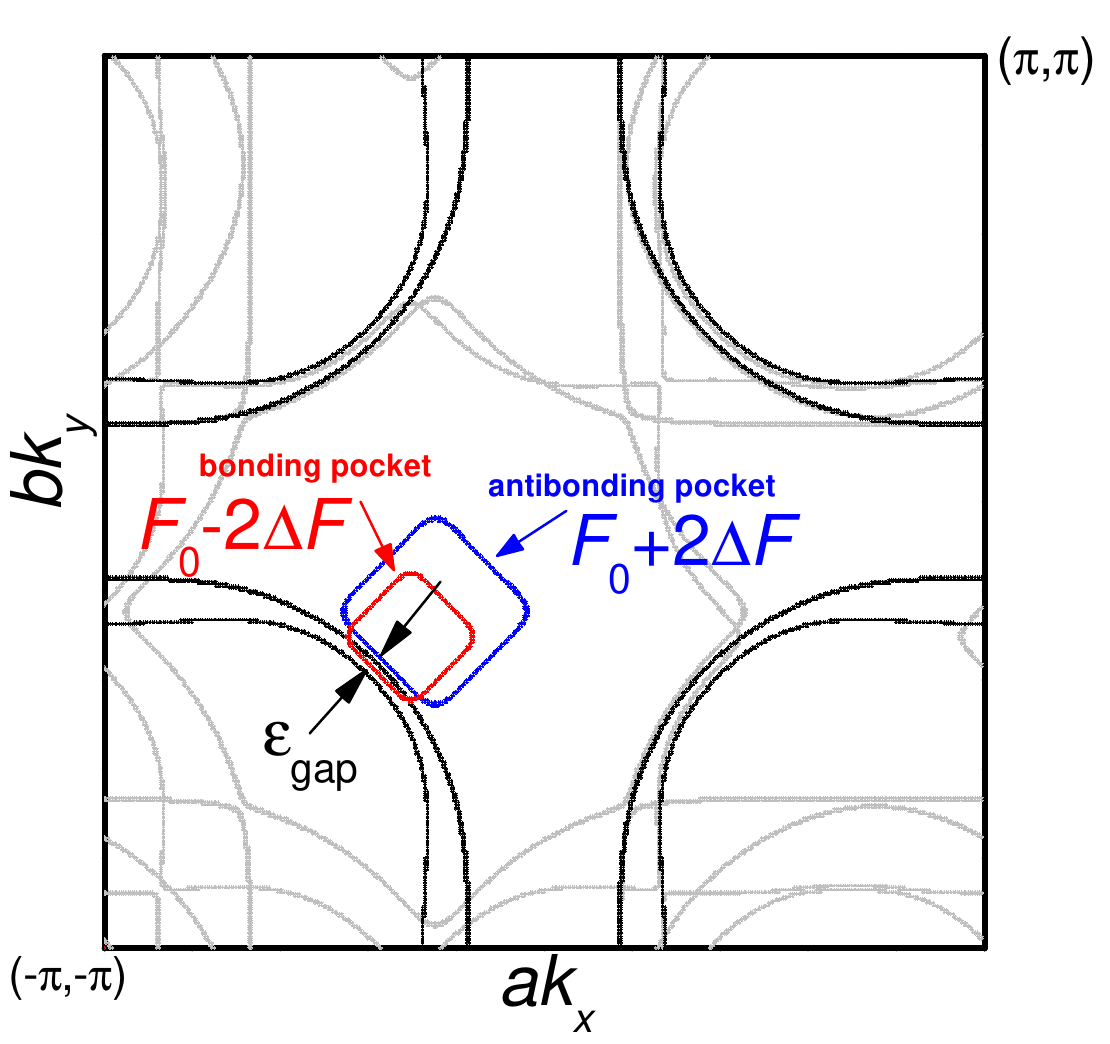}
\caption{Bonding (red) and antibonding (blue) nodal electron Fermi surface pockets obtained using Eq.~(\ref{matrix2}) for $V_0/t=$~0.7, depicted in the extended Brillouin zone representation. Black lines indicate the unreconstructed Fermi surface shown in Fig.~\ref{Fermisurface}(b). Grey lines represent the remaining unnested portions of Fermi surface resulting from the inclusion of only the lowest order terms in Eq.~(\ref{matrix2}). The grey lines become reconstructed into multiple repetitions of the pockets throughout the extended Brillouin zone on including more terms in the Hamiltonian~\cite{harrison2}. A schematic representation of the complete reconstructed Fermi surface is produced by translating the closed pockets by $\pm{\bf Q}_a^{\rm B}$, $\pm{\bf Q}_b^{\rm B}$, $\pm{\bf Q}_a^{\rm AB}$ and $\pm{\bf Q}_b^{\rm AB}$ and all multiples and combinations thereof (not shown for clarity).}
\label{pockets}
\end{figure}

We procede to obtain the Fermi surface for a given strength ($V_0$) of the potential by determining the minimum in the electronic density of states (DOS) for the bonding and antibonding bands as a function of ${\bf Q}_{a,b}^{\rm B}$ and ${\bf Q}_{a,b}^{\rm AB}$. We utilize the fact that the DOS is further minimized by considering momentum-dependent couplings of the form~\cite{harrison2}
\begin{eqnarray}\label{nestingselective}
V_a({\bf k})^{\rm B,AB}=V_0\tfrac{1}{1-r^{\rm B,AB}}(1-r^{\rm B,AB}\cos bk_y)~\nonumber\\V_b({\bf k})^{\rm B,AB}=V_0\tfrac{1}{1-r^{\rm B,AB}}(1-r^{\rm B,AB}\cos ak_x).
\end{eqnarray}
In real-space, $r$ corresponds to a modulation of bond strength, for which there is evidence in recent experiments~\cite{achkar1}. The value of $r$ is tuned so as to suppress the energetically unfavourable coupling between translated states whose group velocities point in the same direction. By adjusting  $V_0$, $t_{\perp0}$, $r^{\rm B}$ and $r^{\rm AB}$ (where B and AB refer to the bonding and antibonding bands), we are then able to tune the locations in $\lambda^{\rm B}$ and $\lambda^{\rm AB}$ of the minima in the DOS.

\begin{figure}
\centering 
\includegraphics*[width=.46\textwidth]{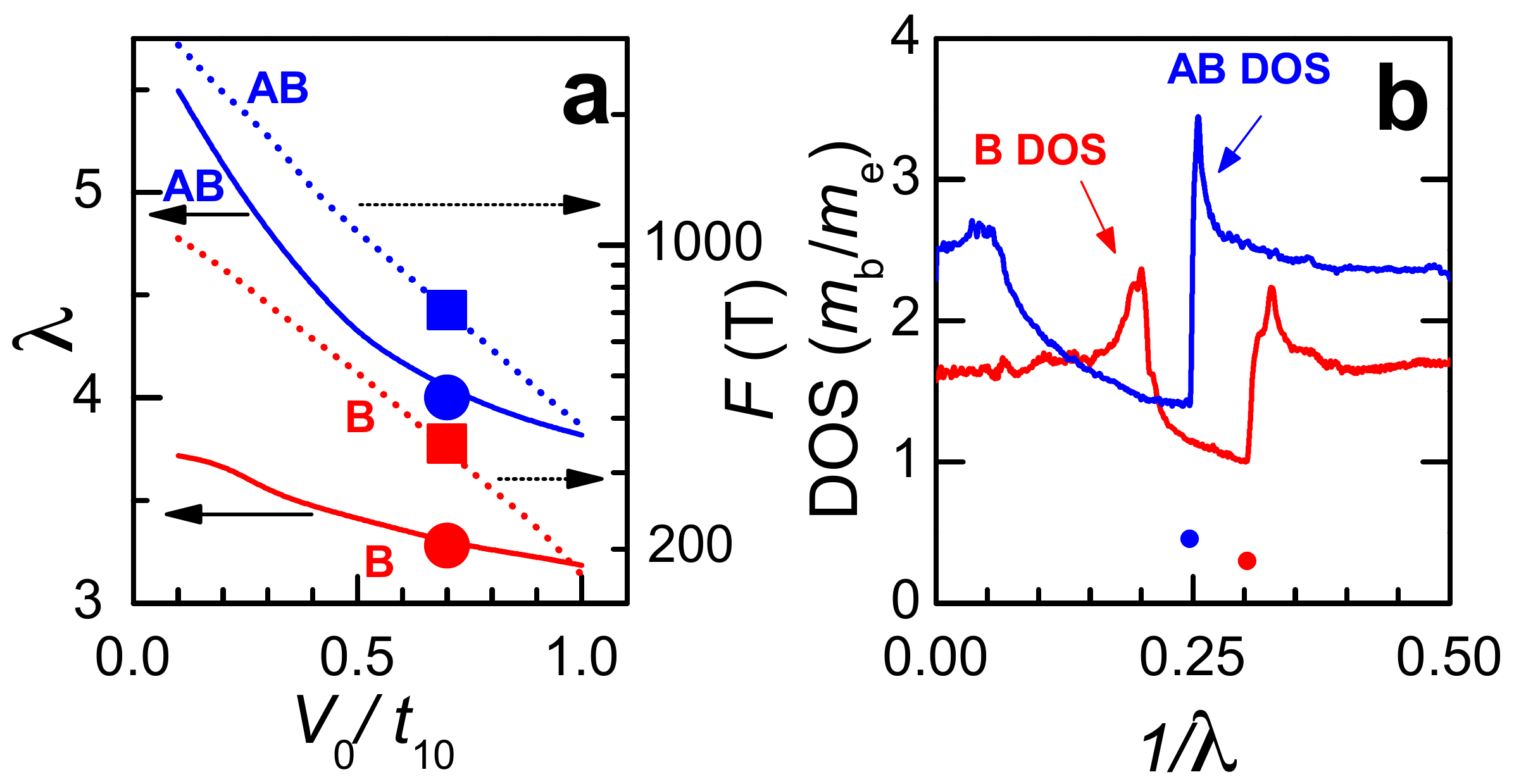}
\caption{(a) The solid lines (left-hand axis) show the optimal values of $\lambda^{\rm B}$ and $\lambda^{\rm AB}$ at which the electronic density-of-states (DOS) is minimized, which are found to depend on the coupling strength $V_0$. Both the bonding (B) and antibonding (AB) bands are shown. At $V_0/t_{10}=$~0.7, the values of $\lambda$ agree with those obtained from x-ray scattering~\cite{ghiringhelli1} and NMR~\cite{wu1} experiments (filled circles). The dotted line (right-hand axis) shows the corresponding pocket frequencies calculated for the bonding and antibonding bands [related to the pocket area via Onsager's relation $F=(\hbar/2\pi e)A_k$], which also depend on $V_0$. At $V_0/t_{10}=$~0.7, the frequencies coincide with the bilayer-split extremal orbits $F_0\pm2\Delta F$ (filled squares) inferred from quantum oscillations in which the frequencies $F_0$ and $F_0\pm\Delta F$ are observed~\cite{sebastian2}.
(b), Electronic DOS for $V_0/t_{10}=$~0.7 (plotted in units of the band mass $m_{\rm b}$ per band in free electron masses $m_{\rm e}$) at the Fermi energy as a function of $1/\lambda$, revealing a clear gap for each band with a minimum that is used to infer the optimal value of $\lambda$. Red and blue lines correspond to the DOS of the reconstructed bonding and antibonding bands obtained from Eq.~(\ref{matrix2}). Dots indicate the minimum DOS per band within the reconstructed Brillouin zone when all multiples of ${\bf Q}_{a,b}^{\rm B,AB}$ are included in Eq.~(\ref{matrix2})~\cite{footnote1}.}
\label{dos}
\end{figure}

We find that the wavevectors ${\bf Q}_{a,b}^{\rm B}$ and ${\bf Q}_{a,b}^{\rm AB}$ at which the electronic DOS is minimized when $V_0/t_{10}=$~0.7, $t_{\perp0}/t_{10}=$~0.54, $r^{\rm B}\approx$~1.65 and $r^{\rm AB}\approx$~1.35 correspond closely to the wavevectors reported in x-ray scattering and NMR experiments. These {\it same} wavevectors also produce nodal pockets (see Fig~\ref{pockets}) with areas consistent with those found in recent quantum oscillation experiments uncovering a bilayer-split Fermi surface [see Fig.~\ref{dos}(a)]. Figure~\ref{dos}(b) shows the electronic DOS (solid lines) for $V_0/t_{10}=0.7$ plotted as a function of $1/\lambda$ for both the bonding and antibonding bands in units of the effective mass. The deep minima (where the free energy is minimized) are located at  $\lambda_{a,b}^{\rm B}=$~3.30 and $\lambda_{a,b}^{\rm AB}=$~4.05, which are very similar to the values $\lambda_{a,b}=$~3.28 and $\lambda_a=$~4.00 reported in x-ray scattering~\cite{ghiringhelli1,chang2,achkar1} and NMR~\cite{wu1} experiments, respectively, performed on underdoped YBa$_2$Cu$_3$O$_{6+x}$~\cite{footnote1}. 

Using Onsager's relation, $F=(\hbar/2\pi e)A_k$ (where $A_k$ is the Fermi surface orbit cross-section in momentum space), we obtain $F^{\rm B}\approx$~337~T and $F^{\rm AB}\approx$~709~T quantum oscillation frequencies for the model Fermi surface pockets in Fig.~\ref{pockets} when $V_0/t_{10}=$~0.7 (see Fig.~\ref{dos}). The size of the closed orbits is not found to be significantly impacted by the inclusion of higher order terms in the Hamiltonian in the limit $|V_{a,b}({\bf k}^\prime)|<t_{10}$, justifying our neglect of such terms for estimating the orbit areas and their effective masses. The observed frequencies would be further modified due to magnetic breakdown tunnelling, which is expected due to the suppressed gap $\varepsilon_{\rm gap}\approx$~10~meV between bonding and antibonding bands at the nodes in underdoped YBa$_2$Cu$_3$O$_{6+x}$, as seen in photoemission experiments~\cite{fournier1}. The low effective mass of $m_0\approx$~0.34~$m_{\rm e}$ for the prominent frequency $F_0$ from band structure calculations~\cite{andersen1,millis1} due to the hopping parameter $t_{10}=$~380~meV implies a renormalization of the bands (and subsequently $V_0$) by a factor of $\approx$~4.7 in the vicinity of the Fermi energy for consistency with the experimentally measured value $m^\ast\approx$~1.6~$m_{\rm e}$. Such a value of the renormalization yields for the residual corrugation (neglected in the two-dimensional Fermi surface approximation) $(t^2_c/t_{\perp0})/$4.7 $\approx$~2.5~meV, which is comparable in magnitude to the small interlayer coherence temperature ($\approx$~27~K) reported in interlayer transport experiments~\cite{vignolle2}. Finally, we obtain $V_0=$~0.7~$\times$~380$/$4.7~$\approx$~57~meV for the strength of the charge order parameter, which, on taking $T_0\approx$~135~K~\cite{chang2} for the ordering temperature, yields  $2V_0/k_{\rm B}T_0\approx$~10. While this ratio is large, it lies within the range of values, 6~$<2V_0/k_{\rm B}T_0<$~17, found exprimentally in typical charge-density wave systems~\cite{fournel1,herr1,johnston1,ozaki1,goserich1}.

\begin{figure}
\centering 
\includegraphics*[width=.46\textwidth]{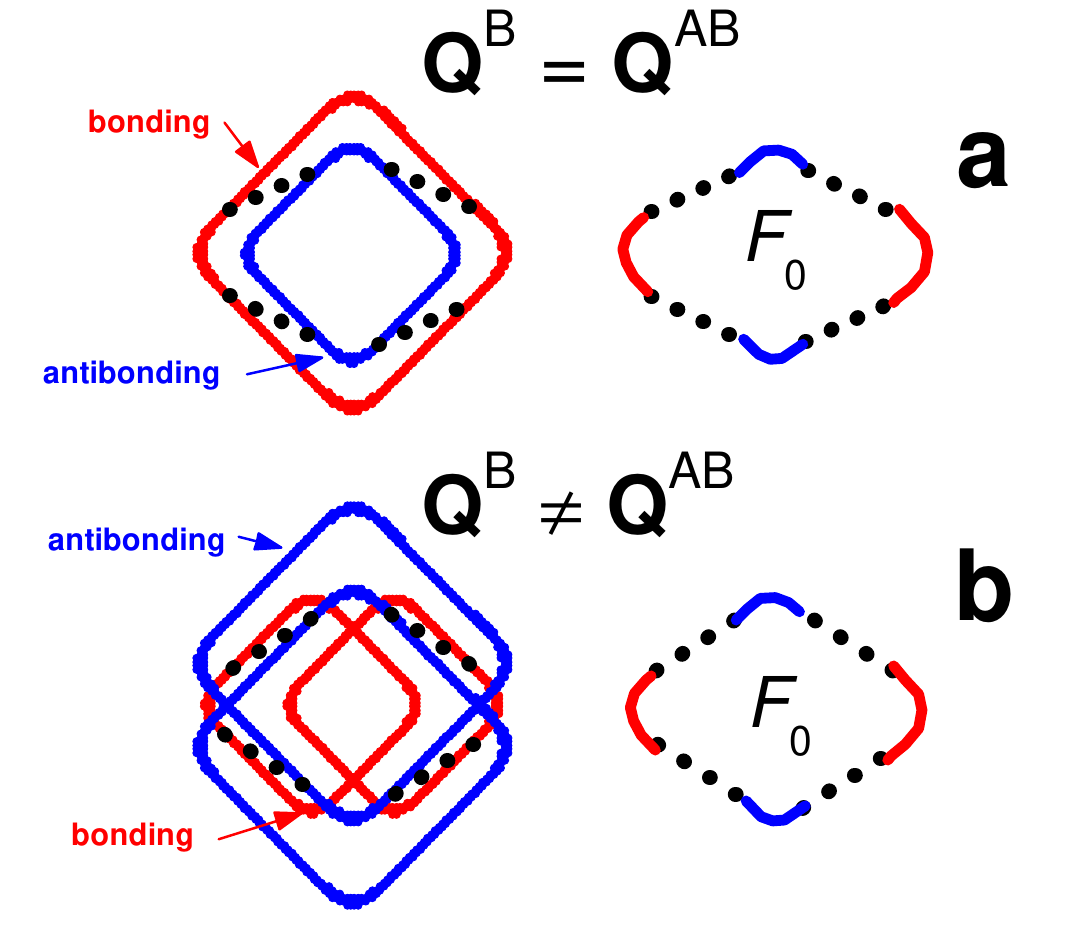}
\caption{(a) Schematic showing how magnetic breakdown tunneling (black dotted line) leads to the observation of the $F_0$ frequency, which dominates the quantum oscillation Fourier spectrum, considering the hypothetical case where the bonding and antibonding charge ordering vectors are the same (i.e. ${\bf Q}^{\rm B}={\bf Q}^{\rm AB}$) after Ref.~\cite{sebastian2}. On the left-hand-side, magnet breakdown tunneling of probability $p^2=\exp(-\varepsilon^2B/\hbar^2\omega^2_{\rm c}F)$~\cite{shoenberg1} (where $\omega_{\rm c}=eB/m^\ast$ is the cyclotron frequency) across the nodal gap $\varepsilon_{\rm gap}$ separating bonding and antibonding Fermi surfaces (indicated in red and blue respectively) is seen to occur at four equivalent points by symmetry  around each of the reconstructed pockets. The resulting magnetic breakdown orbit is indicated on the right-hand-side. (b) On considering the case relevant to the present paper in which ${\bf Q}^{\rm B}\neq{\bf Q}^{\rm AB}$, several pockets with relative momentum space translations of $\pm({\bf Q}^{\rm B}-{\bf Q}^{\rm AB})$ are incorporated into the magnetic breakdown path (left-hand-side schematic). For clarity, only those pockets directly involved in magnetic breakdown tunneling are shown (the complete Fermi surface would consist of the repeated translation of these pockets by all possible combinations of ${\bf Q}_a^{\rm B}$, ${\bf Q}_b^{\rm B}$, ${\bf Q}_a^{\rm AB}$ and ${\bf Q}_b^{\rm AB}$). The magnetic breakdown orbit arising from the four equivalent symmetry-related points of magnetic breakdown tunneling (neglecting smaller gaps) in this case has the same frequency $F_0$ (right-hand-side) as that in (a). }
\label{magneticbreakdown}
\end{figure}

The magnetic breakdown resulting from the small gap $\varepsilon_{\rm gap}$ between bonding and antibonding bands at the nodes (see Fig.~\ref{magneticbreakdown}) is expected to occur at four points around each pocket (in the repeated Brillouin zone), giving rise to a sequence of five frequencies $F_0+m\Delta F$ (where $m=$~$-$2, $-$1, 0, 1 and 2) in which the original bilayer-split frequencies correspond to $F^{\rm B}=F_0-2\Delta F$ and $F^{\rm AB}=F_0+2\Delta F$~\cite{sebastian2} (see Fig.~\ref{pockets}). A small magnetic breakdown gap ($\varepsilon_{\rm gap}$) would further cause the central frequency $F_0$ (depicted in Fig.~\ref{magneticbreakdown}) to dominate~\cite{sebastian2} with weaker adjacent features at $F_0-\Delta F$ and $F_0+\Delta F$~\cite{sebastian2}, while the spectral features at $F_0\pm2\Delta F$ are expected to be further suppressed in amplitude~\cite{sebastian2}. The calculated values $F_0\approx$~519~T and $\Delta F\approx$~95~T compare favorably with those $F_0\approx$~532~T and $\Delta F\approx$~90~T found experimentally~\cite{sebastian2}. 

We turn to the question of why the wavevectors for the bonding and antibonding bands might be detected in different experiments~\cite{ghiringhelli1,wu1} and consider whether details of the Fermi surface reconstruction could provide potential clues. X-ray scattering experiments are expected to be sensitive to the momentum-dependent susceptibility~\cite{ghiringhelli1,chang2,achkar1}, which is largest for the sections of the Fermi surface that are better nested and most completely gapped once Fermi surface reconstruction takes place. We find the electronic DOS to be lower for the bonding band (reflecting its flatter topology), which could potentially account for the observation of $\lambda^{\rm B}$ in x-ray scattering experiments~\cite{ghiringhelli1,chang2,achkar1}. Further experiments are required to determine whether a secondary feature can be seen with x-rays corresponding to the antibonding band at  $\lambda_{a,b}^{\rm AB}\approx$~4; possibly induced by the application of a magnetic field at low temperatures~\cite{FICDW}. NMR, by contrast, is expected to yield a splitting when the charge modulations become commensurate with (or locks-in to) the underlying crystalline lattice~\cite{wu1}; owing to its proximity to an integer value, such a possibility is more likely for $\lambda^{\rm AB}$.

We also note that whereas the formation of a CuO$_2$ plane-centered charge modulation involving the bonding band~\cite{chang2} is mostly decoupled from oxygen ordering within the chains, as indicated by recent x-ray scattering experiments~\cite{achkar1}, the opposite may be true for a charge modulation involving the antibonding band owing to a greater overlap of its wave function with the chains (the chains themselves being prone to different forms of order~\cite{zabolotny1,andersen3}). Local interactions with the oxygen order could cause the latter to become more susceptible to a lock-in transition or to differences in the strength of the modulation along the $a$ and $b$ lattice directions, potentially revealed in NMR experiments~\cite{wu1}.

In summary, we have shown that the different wavevectors reported in x-ray scattering and NMR experiments and the frequencies observed in quantum oscillation experiments~\cite{sebastian2} can be consistently explained by biaxial charge ordering involving different nesting vectors for the bonding and antibonding bands in the bilayer cuprate YBa$_2$Cu$_3$O$_{6+x}$~\cite{sebastian2,andersen1,garcia1}. We note that the negative electrical transport coefficients, size of the high magnetic field electronic heat capacity and chemical potential oscillations are similar to those expected in previously considered biaxial charge ordering models in which the effects of bilayer coupling were neglected~\cite{harrison1,harrison2,sebastian4}. The high purity of YBa$_2$Cu$_3$O$_{6+x}$ samples together with their suitability for both quantum oscillation and x-ray scattering experiments make these materials a model system for understanding the interplay between charge ordering and superconductivity in the cuprates.

This work is supported by the Royal Society, King's College (Cambridge University), US Department of Energy BES ``Science at 100 T," the National Science Foundation and the State of Florida.


\begin{thebibliography}{99}
\bibitem{heatnote} Assuming that the electronic contribution to the heat capacity approaches the normal state value asymptotically as the upper critical field $H_{\rm c2}$ is approached~\cite{kusunose1}, where $\mu_0H_{\rm c2}\lesssim$~100~T in underdoped YBa$_2$Cu$_3$O$_{6+x}$.
\bibitem{choice} This choice approximately reproduces the form of the large planar Fermi surface sections found in several different calculations of the electronic band structure~\cite{andersen1,elfimov1,carrington1} that is broadly consistent with photoemission experiments~\cite{hossain1,fournier1}. Variations in the hopping parameters will affect the values of $\lambda$ and $F_0$ in the model.
\end{thebibliography}
\end{document}